\documentclass[twocolumn,showpacs,prl,superscriptaddress]{revtex4-2}

\usepackage{amsmath}
\usepackage{tabularx}
\usepackage{graphicx}
\usepackage{color}
\usepackage{times}

\graphicspath{{figures/}{data/}}

\begin{document}

\title{Non-universality for Crossword Puzzle Percolation}

\author{Alexander K. Hartmann}
\affiliation{Institut f\"ur Physik, Universit\"at Oldenburg, D-26111
	     Oldenburg, Germany}

\begin{abstract}
A percolation model inspired by crossword puzzle
games is introduced. A game proceeds by solving words, which are segments
of sites in a two-dimensional lattice.
As test case, the \emph{iid} variant allows for independently
occupying sites with letters, only the percolation criterion depends on the
existence
of solved words. For the \emph{game} variant, inspired by real
crossword puzzles, 
it becomes more likely to solve crossing words which share sites
with the already solved words. In this way avalanches of solved words
may occur. Both model variants exhibit a percolation transition
as function of the a-priori site or word solving probability,
respectively. The \emph{iid} variant
is in the universality class of standard two-dimensional percolation.
The \emph{game} variant exhibits a non-universal critical exponent $\nu$
of the correlation length. The actual value of $\nu$ depends on the function
which controls how much solved words accelerate the solved of crossing
words.
\end{abstract}


\maketitle

\section*{Introduction}
\label{sec:intro}

When a statistical physicist looks at a partial solved crossword puzzle,
she or he sees immediately a percolation problem: Is there a spanning path
consisting of fully solved words? Cross word puzzles have been investigated
in the mathematical literature so far with respect to their graph
network structure \cite{mcsweeney2016,cote2021}.

Percolation problems \cite{flory1941,broadbent1957,stauffer2003,saberi2015} are
investigating the conditions for the existence
of system-wide connect components, allowing for
the transport of matter, information or currents. Percolation
is ubiquitous  
in all fields of sciences like physics, mathematics, computer science,
social sciences or biology \cite{stauffer2003,sahimi2023}. Typically,
there are one or more external parameters, like
the density of relevant objects in the system, which control whether the system
is in the percolating phase or in the non-percolating phase.
The transition between the two phases occurs often as a second-order phase
transition, described by critical exponents, e.g., the critical
exponents $\nu$ for the correlation length. Interestingly, the
percolation transition is highly universal, i.e. for different systems 
of the same dimensionality, the critical exponents are the same, independent
of the physical details, in particular $\nu=4/3$ in two dimensions.

This is also true not only for different lattice structures,
but also when changing the geometric properties of the
system's constituents. In particular
the percolation transition for independently placed
dimers \cite{harder1986,dolz2005dimer}, string-like objects as rods and
$k$-mers \cite{cornette2003,dolz2005polyatomic,cornette2006,longone2012} or in general 
objects with varying aspect-ratio \cite{mecke2002}  
is described
by the standard critical exponents.

Most of the standard percolation problems
studied in statistical mechanics consist of disorder without correlations, 
although several studies with imposed correlations exist, in particular
with power-law correlations
\cite{schrenk2013,zierenberg2017}. Here indeed a non-universality
can arise.
For long-range power-law correlated order $C(r)\sim r^{-a}$
and very long-range correlations with $a<3/2$, the
exponent $\nu$ is expected to depend on $a$ \cite{schrenk2013}.
But for values $a>3/2$
the standard 2d value $\nu=4/3$
for the critical exponent
of the percolation length is obtained \cite{zierenberg2017}.

The crossword puzzle 
percolation problem studied here is based on a notion of occupancy, which
is based on linear segments of sites, i.e., \emph{words}. This linear property
is similar to the percolation problem for rods,
which, as mentioned, on its own does not lead to a change of the value of $\nu$.
In addition, for the present crossword model correlations 
arise naturally, because words, where some letters are already known
through other solved words, are easier to solve.
As it is shown below, the crossword puzzle percolation belongs to a
different universality class as compared to standard percolation,
and the value critical exponent $\nu$ depends on how much one benefits
from the partial knowledge of as word.

Note that recently, a model for crossword puzzle was introduced
\cite{mcsweeney2016},
by using quenched normal-distributed
\emph{difficulties} for the words. The difficulty is the fraction
of letters of a word which have to be known in order to solve the word. The
normal
distribution allows for negative difficulties, i.e., words which are
always known, and for large positive difficulties such that these words
cannot be solved. The solution of the words with negative difficulties
lead to other words with positive difficulties to be solved as well,
i.e., creates a dynamic solution process.
The mean and variance of the difficulty distribution determine
whether most words can be solved or not. In the study \cite{mcsweeney2016}
for some parameter 
values a bimodal distribution for the amount of solved words was observed,
which speaks in favor of a first-order phase transition. But no systematic
study with
respect to averaging the grid structure, or the values of the
parameters of the Gaussian distribution or grid sizes was performed.
Thus, in particular the question about the universality could not be addressed.

In the present paper a model is studied which is based
on similar concepts as
Ref.~\cite{mcsweeney2016}, but exhibits higher similarity with
respect to the control parameter as standard
statistical physics percolation models.
Namely, here the a-priori
probability of occupying sites or solving words, depending
on the model variant, is varied.
Also, the present model exhibits a parameter which controls
how much one benefits from the partial knowledge of words. Furthermore,
the present work contains
a comprehensive simulation \cite{practical_guide2015}  study
including finite-size scaling analyses \cite{cardy1988}
which allows one to determine critical exponents like $\nu$,
showing the non-universality of the model.

The paper is organized as follows: First the model is introduced,
for two different variants. Next
the algorithm to determine clusters, based on standard
depth-first search, is outlined. In the main section,
the results for the two variants are shown. Finally, a discussion
is given.

\section*{Model}

The model used here, namely the \emph{game} variant, see below,
is based on the
principle that solving a word makes solving other words wore likely
 \cite{mcsweeney2016}. For a higher flexibility of the present model,
 a word-solution probability function is used instead of fixed
 probabilities.
Still, since solving a word is a stochastic event,
some words with the same probability might be solved while others might not.
The relevant word solution probability
is set at a given initial
probability, which is the main control parameter. Hence it states
the a-priori global fraction of solved words.
Thus, the model is  more in the spirit of a standard
percolation probability and correspondingly
leads to a second-order phase transition.

More formally, a
realization of the problem is given by variables $s({\bf x})\in \{-1,0,1\}$
for sites ${\bf x}$ on a $d$-dimensional lattice of site $L^d$.
 The three different
 values correspond to \emph{black},
 \emph{empty} and \emph{occupied}
 sites, respectively. With respect to a real crossword puzzle,
 an occupied site contains a letter. Black sites cannot contain letters.
 Sites which are not black are also called \emph{white}.
Here, the
simple quadratic, i.e., 2d, case  
with periodic boundary conditions in all directions
is considered, a generalization to other dimensions $d$ is straightforward.

\begin{figure}[ht]
\includegraphics[width=0.45\columnwidth]{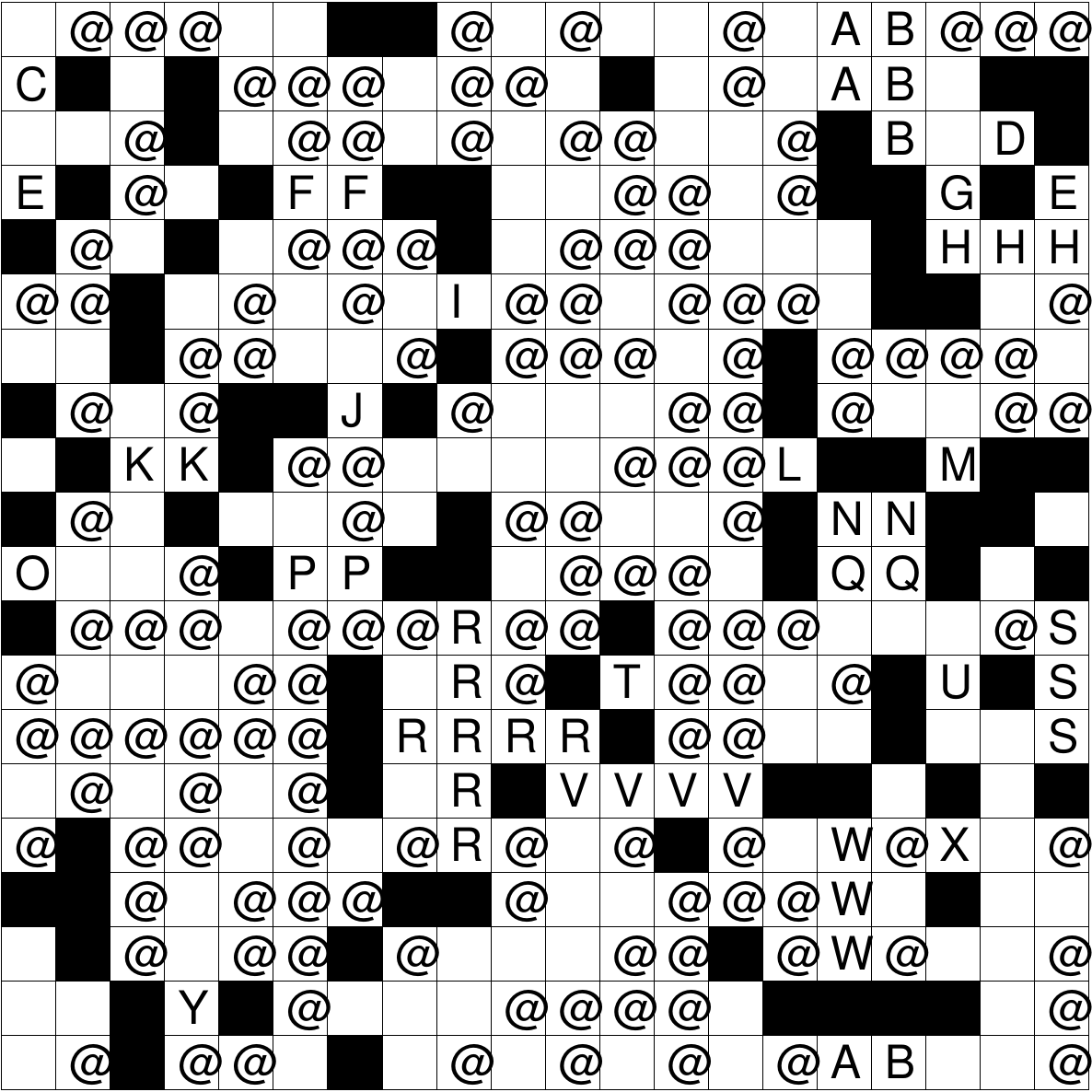}
\hspace*{0.04\columnwidth}
\includegraphics[width=0.45\columnwidth]{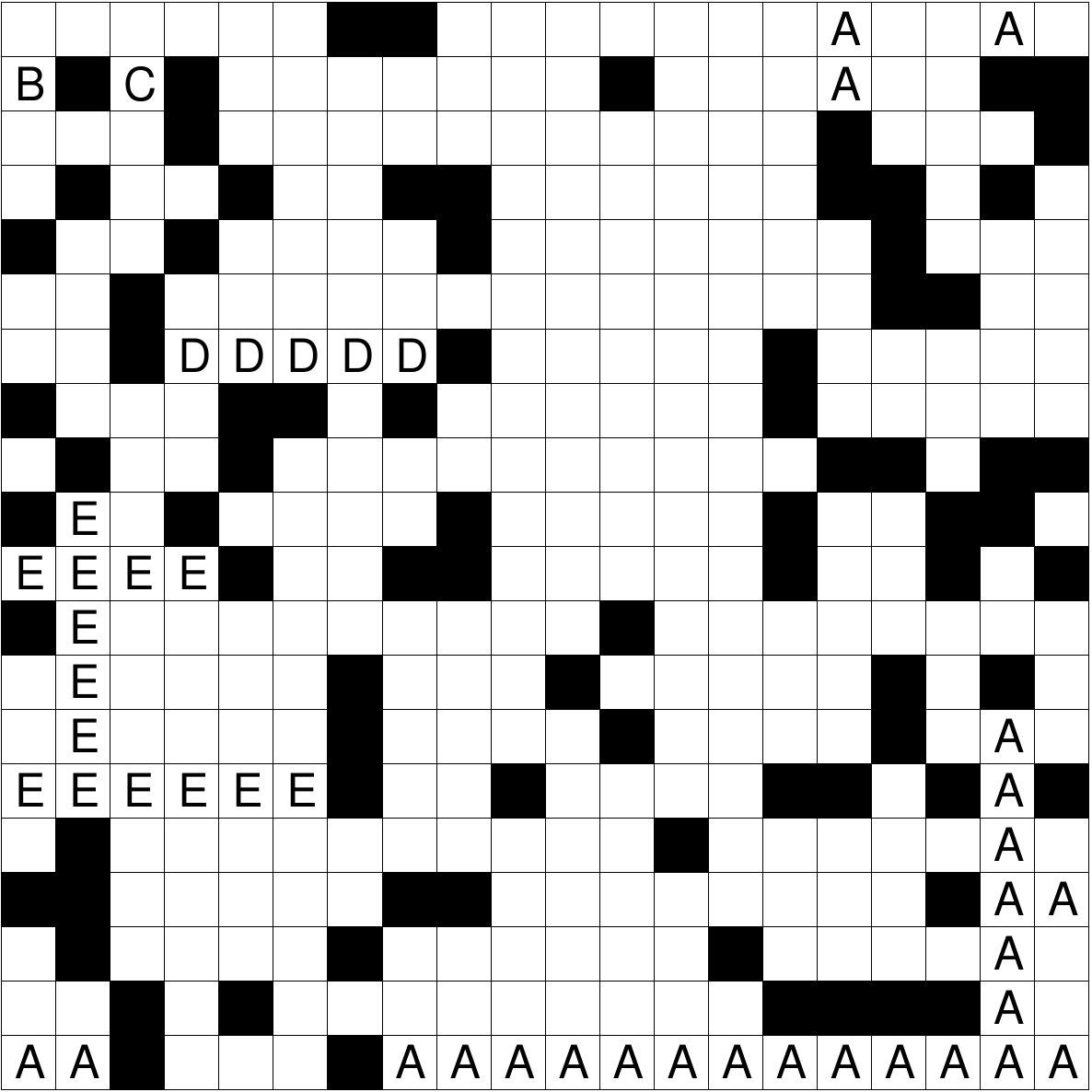}
\caption{\label{fig:samples}
Sample realizations for $L=20$, $p_{\rm b}=0.2$. Black sites are black,
empty sites white, occupied sites contain symbols and letters.
Any segment of sites bordered by black sites is a word.
The letters A,B,\ldots
denote the clusters formed by solved words, while the @ symbol
denotes occupied sites which do not belong to solved words.
In the left an \emph{iid}
sample is shown for $p=0.5$. In the right an \emph{game} sample is
shown for $p_{\rm w}=0.053$, $\omega=1$.
}
\end{figure}

Realizations are generated as follows: First, 
 each site is set to \emph{black} with probability $p_{\rm b}$.
All other sites are white, i.e., empty or occupied with a letter.
Two examples are shown in Fig.~\ref{fig:samples}.
The black sites partition the system into \emph{words}, i.e. segments of
horizontal or vertical white sites bounded by black sites.
Note that the black sites are assigned independently of each other,
such that short words like of length one might occur, in contrast to
typical real-world crossword puzzles. Also very long words might occur:
For the 
special case that a row or column does not contain any black site, the full
row or column is a single word of length $L$, but this occurs only
for very small values of $p_{\rm b}$ and $L$. Technically, for all simulations,
the words of a given realization 
are determined after the black sites have been assigned.

In order to assign the white sites, to be occupied or empty,
two variants are considered, 
the \emph{iid} and the \emph{game} variant. For the \emph{iid} case,
each white site is assigned the state occupied independently 
with the identical probability $p$. With probability $1-p$ the
site is empty. The occupation rule of this variant is similar to standard
percolation and is used for comparison.

The \emph{game} variant mimics the way a human would try to solve
a crossword puzzle.
The a-priori probability $p_{\rm w}$
describes the probability 
to solve a word, i.e., occupy all sites of a word, if no
letters are present so far. Furthermore, if some words
are solved, corresponding to partial knowledge of ``neighbouring'' 
words which
share a letter with the solved word, 
this will increase the probability that a neighboring word can
be solved as well. This is here modeled by the word-solution probability
\begin{equation}
p_{\rm cw}(x)=p_{\rm w}+(1-p_{\rm w})x^{\omega}\,,
\end{equation}
where $x\in [0,1]$ is the fraction of already known letters, i.e., 
occupied sites, of a word. Note that $p_{\rm cw}(1)=1$ which is consistent.
The \emph{benefit exponent} $\omega$, describes how much knowing
some letters helps in solving a word. For $\omega\to\infty$, one
does not benefit much, since $p_{\rm cw}(x)\approx p_{\rm w}$ for
almost all $x\in [0,1]$. For values $\omega<1$ the benefit grows
in particular quickly.

To generate a realization of disorder for the \emph{game} variant, one
draws for each word $w$ a fixed random number $r(w)$. Then, one iterates over
all words and solves a word, i.e. occupies all its sites ${\bf x}$, if 
$r(w)<p_{\rm cw}(x(w))$
where $x(w)$ is the current occupation fraction of word $w$,
which is initially zero for all words. Occupying
the letters for a word leads to an increase of $x(w')$ for words $w'$
which share sites with $w$. This makes it more likely that
$w'$ can be solved as well. The process is iterated again over all words,
until no more additional words are solved.

\section{Algorithm}

To analyze a given realization, let it be the \emph{iid} or the
\emph{game} variant, the following procedure is applied:
First, all \emph{solved} words are determined, i.e., those words where all
its sites are occupied. Now, two words are called \emph{connected} if they
are both solved and if they share one site. In particular, they
run perpendicular to each other on the grid, crossing at the shared site.
Thus, two
word which run parallel next to each other are not connected.
 Now, \emph{clusters} 
of words are 
determined by a standard depth-first search \cite{cormen2001}
as the transitive closure
of connected words. The size $s$ of a cluster is the number of occupied
sites. A realization is considered \emph{spanning} if there is a cluster
which exhibits at least one occupied site in each row or in each column,
i.e. it connects the any row (column) with any other row (column).

\section*{Results}

Simulations were performed for several system sizes between $L=20$
and $L=1000$, for the \emph{iid} and the \emph{game} variant. 
For each realization, determined also by the parameters $p_b$, $p$ 
or $\omega$, $p_w$, an average over a number of realizations
was taken, between $10^4$ for system sizes $L\le 200$ and $2\times 10^3$
for $L=1000$.

\begin{figure}[ht]
\includegraphics[width=0.9\columnwidth]{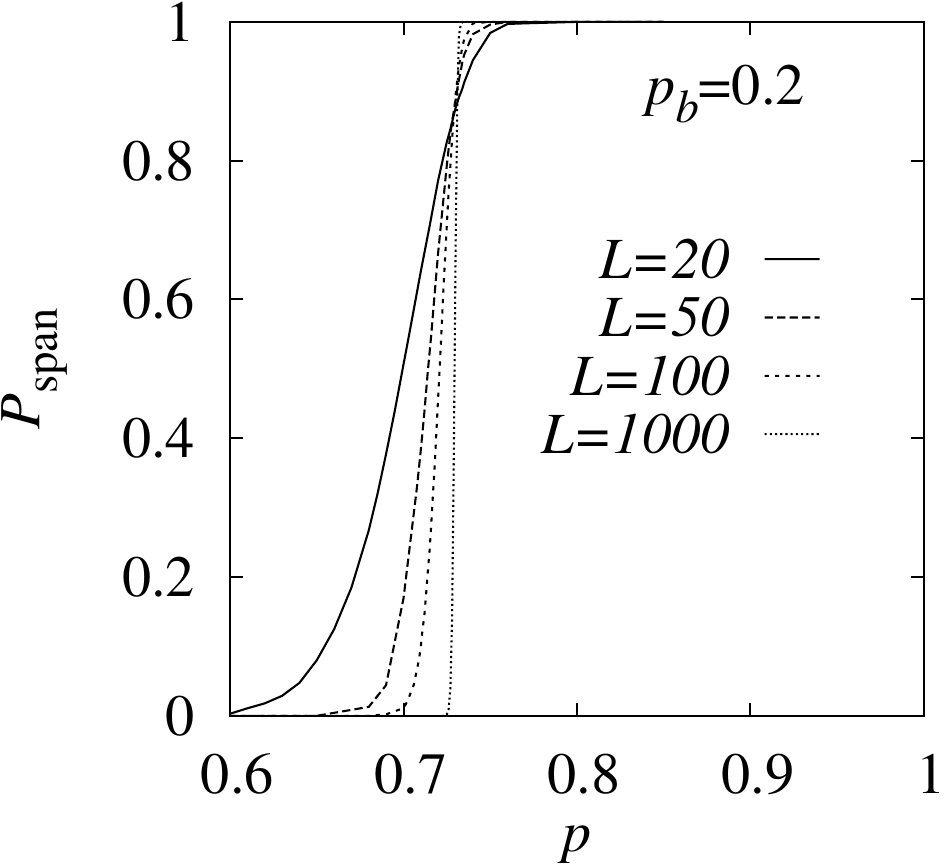}
\caption{\label{fig:p_perc_pe_cw}
Spanning probability $P_{\rm span}$ as function of the \emph{iid}
site occupation probability $p$ for a fraction $p_{\rm b}=0.2$ of black
sites.}
\end{figure}

First, results for the \emph{iid} case are presented. The probability
$P_{\rm span}$ of a spanning cluster is shown in Fig.~\ref{fig:p_perc_pe_cw}
as function of the site-occupation probability $p$, for the case of a fraction
$p_b=0.2$ of black sites. A clear increase of $P_{\rm span}$ near $p=0.73$ is
visible, where the curves of different sizes cross. This is a clear indication
for a percolation phase transition. 

\begin{figure}[ht]
\includegraphics[width=0.9\columnwidth]{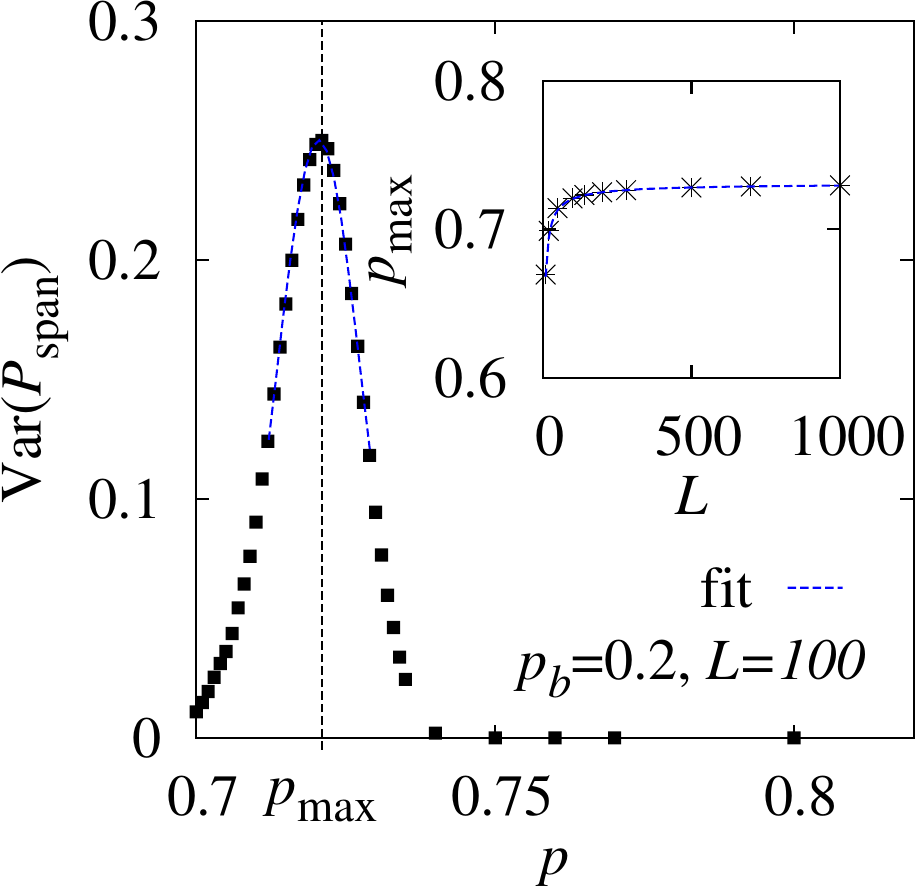}
\caption{\label{fig:var_p_perc}
Variance of the spanning probability $P_{\rm span}$ as function of the 
\emph{iid}
site occupation probability $p$ for $p_{\rm b}=0.2, L=100$. 
Near the peak variance, a Gaussian is fitted to obtain
the peak position $p_{\max}$. The dependence of $p_{\max}$ as function
of system size $L$ is shown in the inset and allows for an extrapolation
to $L\to\infty$ by using Eq.~(\ref{eq:power}).}
\end{figure}

A simple quantitative finite-size analysis of the phase transition is 
possible by considering the variance of the spanning probability
which is simply given by Var$(P_{\rm span})=P_{\rm span}(1-P_{\rm span})$, 
which is shown in Fig.~\ref{fig:var_p_perc} 
as an example for size $L=100$. An estimate of the
finite-size transition point is given by the maximum $p_{\max}$ 
of the variance, where
the sample-to-sample fluctuations are largest. Note that this
also corresponds to $P_{\rm span}=1/2$. The values of $p_{\max}$ were
estimated by fitting Gaussians near the peak.

To extrapolate to the infinite system size, the peak positions are fitted
to the standard finite-size scaling form, well known for standard
percolation \cite{stauffer2003,newman2000} and other phase transitions
\cite{cardy1988}
\begin{equation}
\label{eq:power}
p_{\max}(L)=p_{\rm c}+aL^{-1/\nu}\,,
\end{equation}
where $p_c$ is the critical point and $\nu$ the critical exponent which
describes the divergence of the correlation length. For $p_b=0.2$
the fit for $L>140$ results  in a value of $\nu=1.31(4)$ which 
is compatible with the exponent $\nu=4/3$ for standard 2d percolation.
The resulting critical value is $p_c=0.7313(1)$. 

\begin{figure}[ht]
\includegraphics[width=0.9\columnwidth]{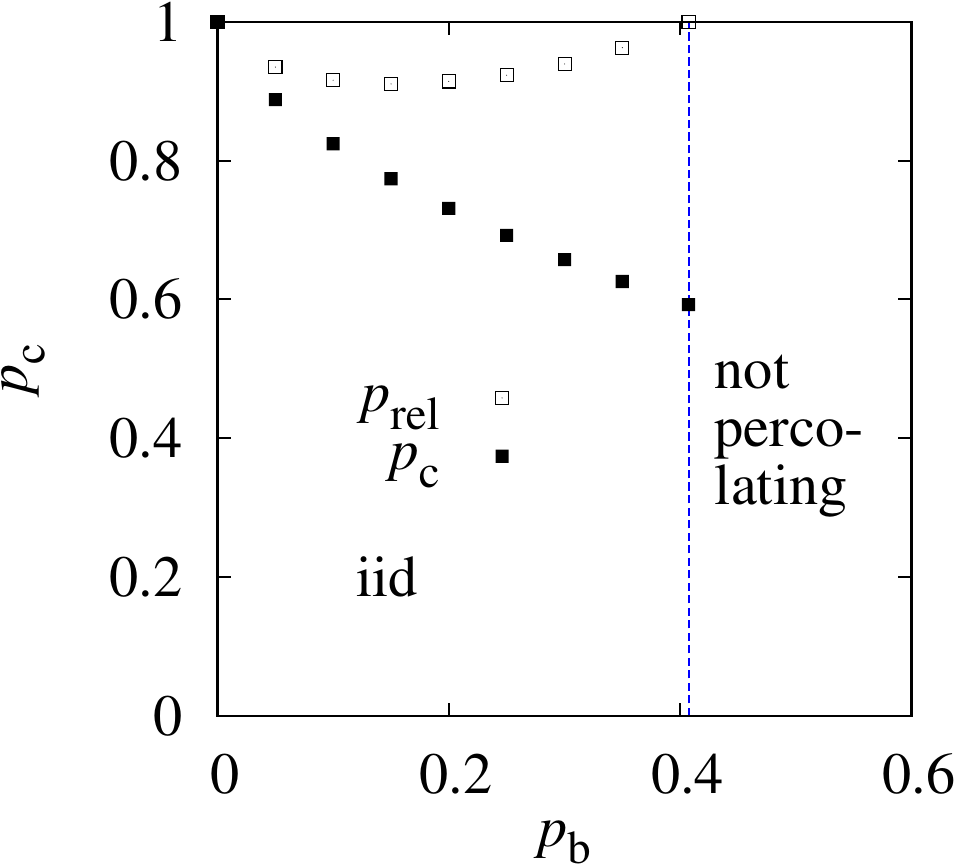}
\caption{\label{fig:p_perc_pe}
Phase diagram if the \emph{iid} case, i.e., critical threshold 
$p_{\rm c}$, together with $p_{\rm rel}=p_{\rm c}/(1-p_{\rm b})$, 
as function of the fraction
$p_{\rm b}$ of black sites. 
}
\end{figure}

The behavior of the critical point $p_c$ as a function of the fraction 
of black sites is shown in Fig.~\ref{fig:p_perc_pe}. For
the border case of no black sites $p_{\rm b=0}$, all words are occupying
a full row
or column respectively.
Here the system behaves
essentially one-dimensional.
Therefore, all sites of a row or column
have to be occupied
for a solved word, i.e., $p=1$.  This is shown in the upper left
corner of Fig.~\ref{fig:p_perc_pe}.

On the other hand, when the fraction $1-p_{\rm b}$ of white
sites reaches the percolation
threshold $p_{\rm c}^{(d=2)}\approx 0.59274621$ \cite{newman2000} of standard
percolation, all white sites have to be occupied for a spanning
path, i.e., $p=p_{\rm c}$. For $p_{\rm b} \ge 1- p_{\rm c}^{(d=2)}$ the
system cannot percolate for any value of $p$. Note that $p$ is measured
as fraction of all lattice sites. One can also measure the fraction
of occupied sites among the white sites, i.e., the relative fraction of
occupied sites. The corresponding
critical fraction is $p_{\rm rel}=p_{\rm c}/(1-p_{\rm b})$ which is also
shown in Fig.~\ref{fig:p_perc_pe}. Interestingly, 
since trivially $p_{\rm rel}(p_{\rm b}=0)=1$
and $p_{\rm rel}(p_{\rm b}=1-p_{\rm c}^{(d=2)})=1$, the behavior of 
$p_{\rm rel}(p_{\rm b})$ is non-monotonous and exhibits a minimum. According to 
the figure this is located near $p_{\rm b}\approx 0.15$.

Next, the results for \emph{game} variant are shown, restricted
to the case $p_{\rm b}=0.2$.
First, the correlations of the disorder correlations are 
quantified via analyzing the density-density correlation function
\begin{equation}
C(r)=\frac{[ s({\bf x})s({\bf x+r}) ]_{0/1} - [ s({\bf x})]_{0/1}^2}
{[ s({\bf x})^2]_{0/1} - [ s({\bf x})]_{0/1}^2} 
\end{equation}
where $r=|{\rm r}|$ amd ${\rm r}$ is for simplicity along the $x$
direction. 
The average $[\ldots ]_{0/1}$ is  
over the disorder realizations and over white sites only because
the location of black sites is not correlated anyway.  
Note that  $[ s({\bf x})^2]_{0/1}=[ s({\bf x})]_{0/1}$
is just the resulting density of occupied sites among the white sites.

\begin{figure}[ht]
\includegraphics[width=0.9\columnwidth]{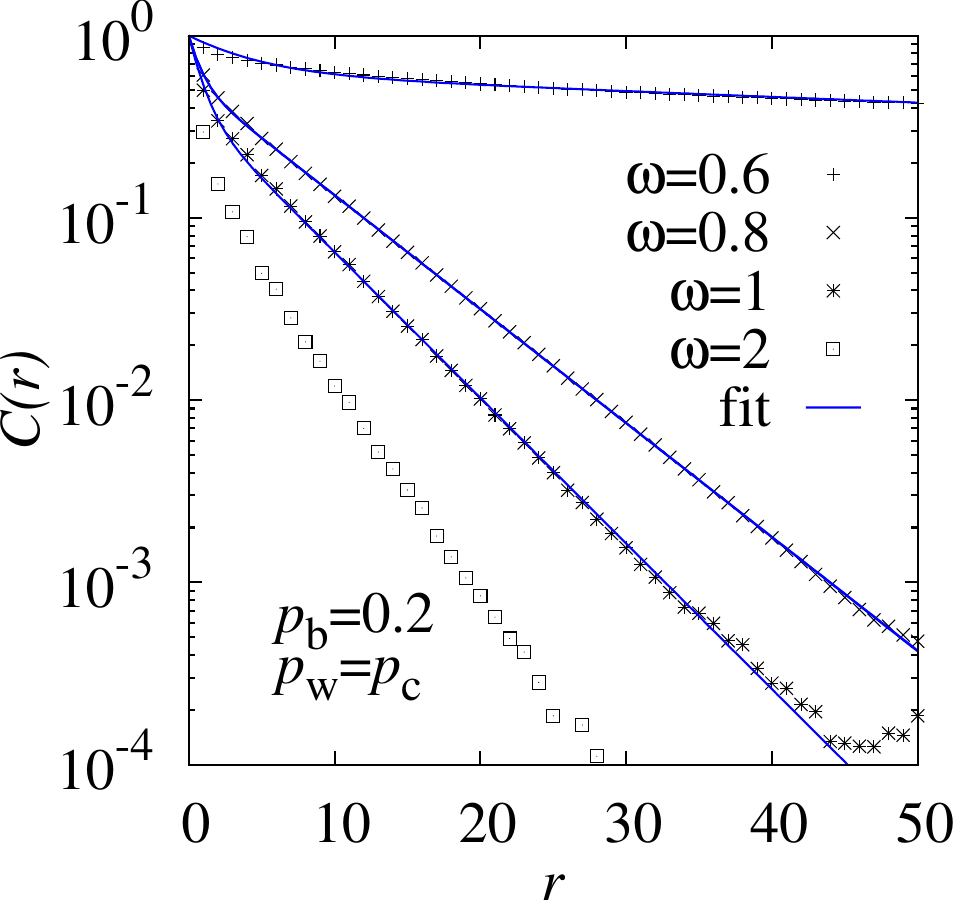}
\caption{\label{fig:correlations_word}
Density-density correlations of the \emph{game} realizations for the considered
values of $\omega$ for  the critical value $p_{\rm w}=p_c$, respectively 
The lines show fits to a sums of two exponentials.}
\end{figure}

In Fig.~\ref{fig:correlations_word} the correlations are shown near
the critical points as determined below, together
with fits to sums of two exponentials $e^{-r/\lambda_1 }+e^{- r/\lambda_2}$
with $\lambda_1>\lambda_2$. For all
values of $\omega$, clear exponential decreases are 
visible. The dominating length scales $\lambda_1$ as
obtain from the fit are for $\omega\in[ 0.8,2]$ small,
i.e. $\lambda_1\le7$. For $\omega=0.6$
a much longer scale $\lambda_1=137(4)$ is obtained. Note that for
the latter case below $p_c=0$ is estimated, so the simulations
were performed at the values $p_w=0.00004$ close to $p_c$.
For larger distances $r$, depending on $\omega$,
the correlations fluctuate around zero
due to the finite statistics. Thus,
the correlations are exponentially decreasing, i.e., short-ranged,
 and one could expect that, in particular for 
$\omega\ge 0.8$, they have no
influence on larger length scales such that the critical
behavior of standard percolation is obtained.

\begin{figure}[ht]
\includegraphics[width=0.9\columnwidth]{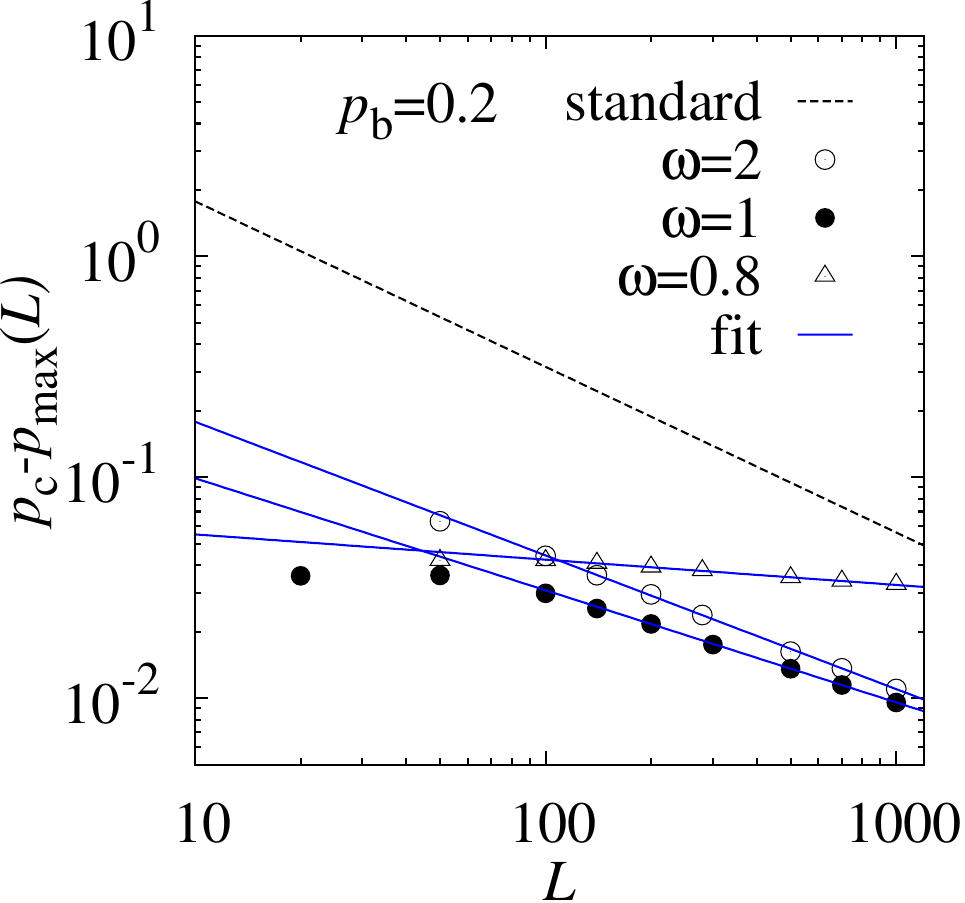}
\caption{\label{fig:peak_L_word}
Approach of the finite-size critical points $p_{\max}$ for the \emph{game} 
variant to the one extrapolated
by Eq.~(\ref{eq:power}), for various values of $\omega$. The slope in
the log-log plot corresponds to the critical exponent $-1/\nu$ and
shows the non-universality of the model.
}
\end{figure}

But this is actually not the case. The spanning probability $P_{\rm span}$
as function of the word probability $p_{\rm w}$ looks qualitatively similar
to the \emph{iid} case, thus it is not shown here.
Although for each realization avalanches of
solved words occur, the function $P_{\rm span}(p_{\rm w})$ looks very smooth, no
indication of a first-order step-like behavior is visible.
One can again fit Gaussians
near the peaks of the variance Var$(P_{\rm span})$ to obtain finite-size
critical points $p_{\max}(L)$. Fitting to  Eq.~(\ref{eq:power}) yields
extrapolated critical points $p_{\rm c}$ and critical exponents $\nu$
of the correlation length. In Fig.~\ref{fig:peak_L_word} the results
for $p_{\max}(L)-p_{\rm c}$ are shown as function of $L$.
For $L\ge 100$ clear power laws are visible,
but with different exponents $-1/\nu$ as compared to the standard percolation
value $-3/4$. For $\omega\ge 0.8$ here always
larger critical exponents $\nu$ are found, see Tab.~\ref{tab:critical}.
This shows that the crossword percolation is
in a different universality class than standard percolation.
Note that with increasing value
of $\omega$, the critical exponent $\nu$ moves towards the standard value.
This makes sense, because the benefit becomes smaller and smaller
when $\omega$ increases, such that the percolation problem
of independent rods should be obtained, which is known to be in the standard
universality
class \cite{cornette2003,dolz2005polyatomic,cornette2006,longone2012}
also for polydisperse systems \cite{mecke2002}.

The table shows that the critical point $p_{\rm c}$ 
decreases, as the benefit exponent $\omega$
decreases, since the growing benefit means that less words have to
be solved without the help of some letters.
Interestingly, for a small values of $\omega=0.6$, a critical value 
which is compatible with $p_{\rm c}=0$
is obtained. Thus, the benefit from solving words partially is so large
that an infinite small a-prior solution probability is sufficient to solve
a full puzzle. Also, 
an even more non-standard value of the critical exponent $\nu$ is found.
In particular 
the $L\to\infty$ approach to the critical point is now, not from below but
from above (not shown). This is natural, because no 
probabilities $p<p_c=0$ exist,
in contrast to larger values of $\omega$ or the 
\emph{iid} case, where the $L\to\infty$ approach is from below as visible
in Figs.~\ref{fig:var_p_perc} and \ref{fig:peak_L_word}.


\begin{table}
\begin{tabular}{l||r|r|r|r|r}
$\omega$ & 0.6 & 0.8  & 1.0 & 1.5 & 2.0\\ \hline
$p_{\rm c}$ & 0 & 0.037(5) & 0.0948(2) & 0.204(1) & 0.264(1) \\
$\nu$ &  0.56(3) & 3.0(10) & 1.96(2) & 1.78(9) & 1.65(11) \\
$\beta$ & -- & 0.05(2) & 0.09(2) & 0.20(2) & 0.19(2) \\
$\tau$ & -- & 2.06(5) & 2.05(5) & 2.07(5) & 2.05(5) \\
\end{tabular}
\caption{\label{tab:critical}
Values of the critical point $p_{\rm c}$ and the critical exponents
$\nu$ 
for the \emph{game} crossword percolation, for various values of
the benefit exponent $\omega$.} 
\end{table}

\begin{figure}[ht]
\includegraphics[width=0.9\columnwidth]{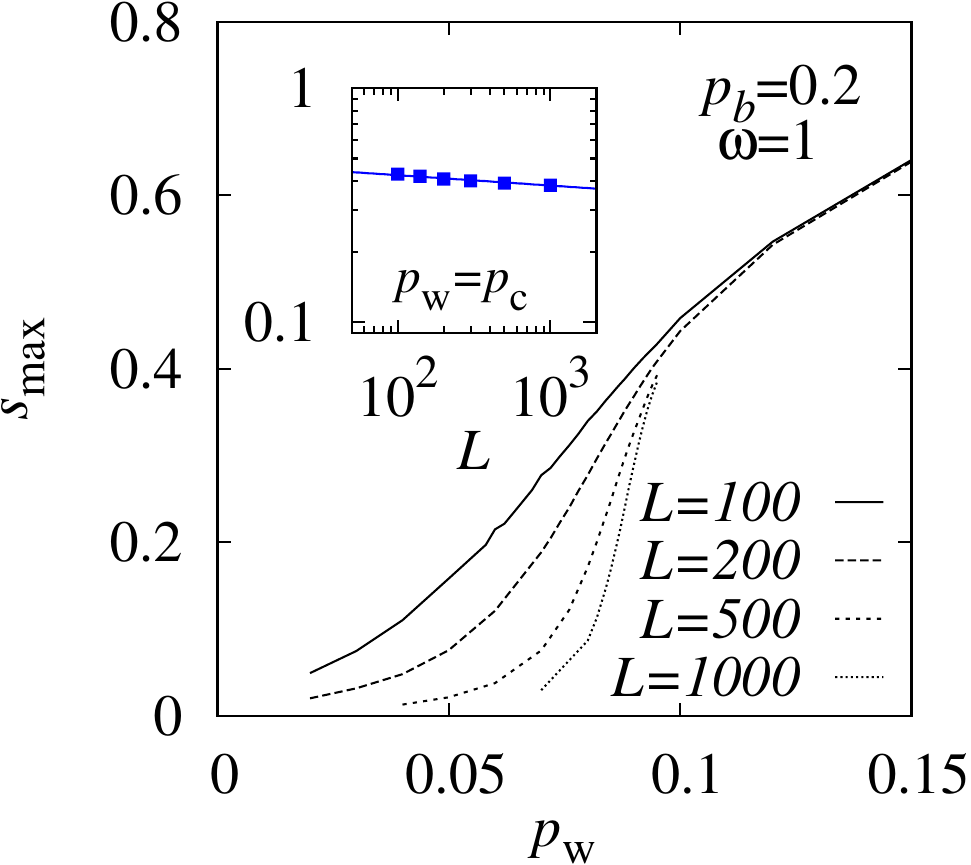}
\caption{\label{fig:smax_pw}
Relative size $s_{\max}$ of the largest component as function of the
word probability $p_{\rm w}$ for the $\omega=1$ \emph{game} case,
for several system sizes $L$. The
inset shows the behavior at the critical point $p_{\rm w}=p_{\rm c}$
as function of system size $L$ together with a fit to a power
law $\sim L^{-\beta/\nu}$.
}
\end{figure}

To gain further insight, the order parameter, i.e., the average size $s_{\max}$
of the larges component as function of $p_{\rm w}$ was studied. For all
values of $\omega$, a smooth behavior was found, see Fig.~\ref{fig:smax_pw}
for the case $\omega=1$. At
the corresponding critical points,
 power laws $\sim L^{-\beta/\nu}$ were observed, except for $\omega=0.6$
where it is not possible to generate data at the critical point $p_{\rm c}=0$.
  In all
cases, values of $\beta$ as reported in Tab.~\ref{tab:critical} are
rather small, such that the
differences to the standard value $\beta=5/36\approx0.139$ are not too large.
Still, for values $\omega<1$ the value of $\beta$ appears to be 
significantly smaller than the standard value.
 Since the scaling of 
the absolute cluster sizes  also defines via $L^ds_{\max}\sim L^{d_{\rm f}}$
the fractal dimension $d_{\rm f}$, this leads to 
the hyper-scaling relation $\beta/\nu=(d-d_{\rm f})$, i.e., 
the fractal dimension $d_{\rm f}\le d$ is close to 2, in particular for
small values of $\omega$, which means that
 the clusters are almost space filling. 

\begin{figure}[ht]
\includegraphics[width=0.9\columnwidth]{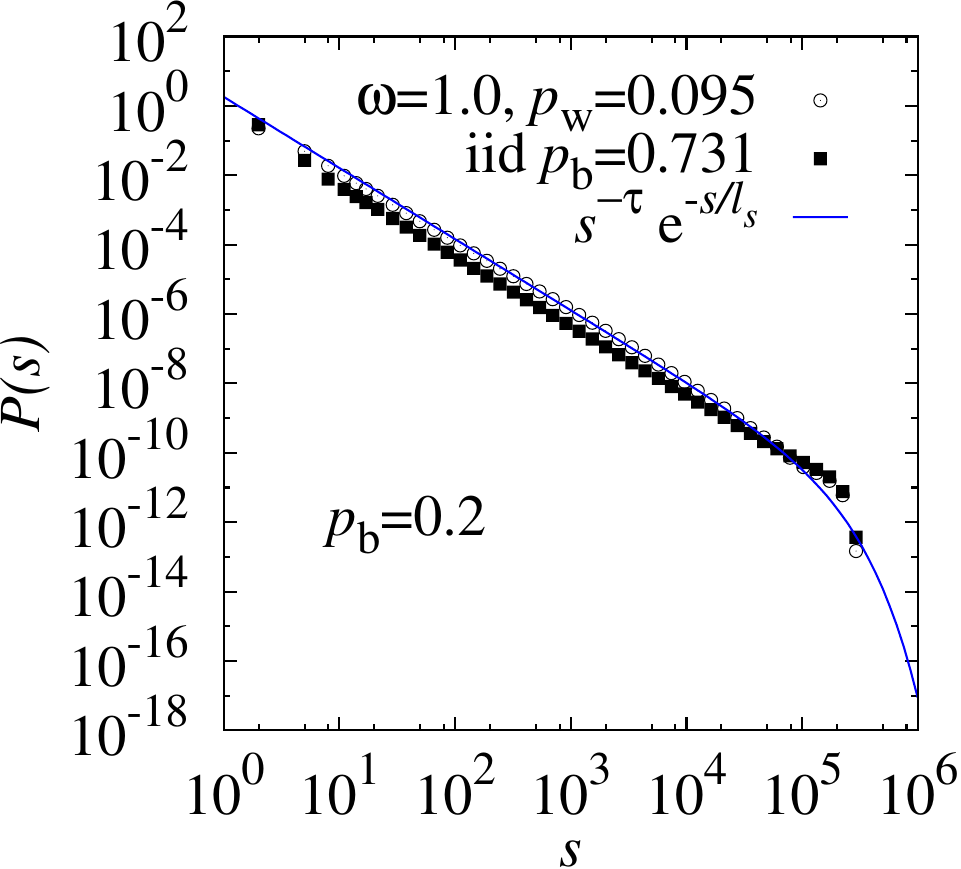}
\caption{\label{fig:distr_size}
Distribution of the sizes of the non-percolating 
clusters at the critical points for the \emph{iid}
case and for the $w=1$ \emph{game} case along with a fit to
 Eq.~(\ref{eq:distr:s}) for the $\omega=1$ case.
}
\end{figure}

Furthermore the
distributions of sizes for the non-percolating clusters at the corresponding 
critical points
were studied,
i.e., the probabilities $P(s)$ that a cluster has size $s$. 
For $\omega=0.6$, the critical estimated point is $p_c=0$, so it is
not possible to obtain $P(s)$ at the critical point.
For $\omega\ge 0.8$, mainly a power-law behaviors $P(s)\sim s^{-\tau}$ 
were observed, each with an exponential cut off due to the finite system
sizes. Two sample results, for the \emph{iid} case and for $\omega=1$
of the \emph{game} case are shown in Fig.\ref{fig:distr_size}.
Fits to 
\begin{equation}
P(s)=Z_s s^{-\tau}e^{-s/l_s}
\label{eq:distr:s}
\end{equation}
 with suitable length scale $l_s$ and normalization $Z_s$ 
were performed. Since the data does 
not exhibit perfect power laws over larger ranges of the size,
the result for $\tau$ depends a bit on the fitting range. Thus,
no very precise results  for the exponent $\tau$ could be obtained, but they
are all scattered near the
standard value $\tau=187/91\approx 2.05$
as quoted in Tab.~\ref{tab:critical}. This holds also for the \emph{iid} case.
From the standard hyper scaling relation $\tau=d/d_{\rm f}+1$ 
this is compatible with the observed high similarity of $d_{\rm f}$ to the standard
value.

\section*{Discussion}

In the present work, crossword-puzzle percolation is introduced,
where letters or words are occupied with independent or neighbor-dependent
probabilities. In the model, letters correspond to sites and words
to segments of sites, bordered by black sites.
 The model comprises properties of several
other non-standard percolation models: Like in rod percolation,
the percolation objects are linear segments. Like in bootstrap
percolation, the occupation of the sites, here words,
influences neighboring site. 
Like for long-range correlated disorder, the resulting  critical
exponents differ from standard percolation, in contrast to bootstrap
and rod-like percolation. Still, the density-density correlations
of the model are only short range. 
Thus, it appears that crossword-puzzle percolation
comprises a new type of universal behavior. One possible
explanation for the non-universality could be that the word-solving
probabilities, although not long-range correlated, change during the
solving process, i.e., solving some word leads to an acceleration
or improvement of solving other words. This dynamic element
bears some similarity to directed percolation \cite{hinrichsen2000},
which is a model
for non-equilibrium dynamic processes. An indeed, directed percolation
is characterized by different values of the critical exponents as
compared to standard percolation.

Also, it is remarkable that for very small values of the benefit exponents,
the numerical results indicate that the percolation transition appears
at infinitely small strength of the disorder. This is also a feature
which is not present in standard percolation.

For further studies it would be certainly of interest to look in more
detail into the model, e.g. by studying fractal properties of clusters,
the backbone and other characteristic quantities.
Certainly,  one should also consider higher
dimensions in order to see whether the non-universality occurs there
as well. Also, one could determine the upper critical dimension, which could
be the same as for standard percolation, or not. 

Also the process of generating a \emph{word} realization could be studied:
Since solving a word leads to an increase of the probability of
solving neighboring words, this leads to further iterations, i.e.,
avalanches of solving
words. Some test runs revealed that these avalanches seem to
be largest right at the critical point, leading to longer
running times there. This phenomenon could be similar 
to critical slowing down as observed for Monte Carlo simulations of
various models like Ising models
\cite{williams1985,swendsen1987,middleton2001b}, or related to avalanches in
systems exhibiting
self-organized criticality \cite{bak1987,turcotte1999}.

Furthermore, so far it is not clear why the model does not belong
to the universality class of standard percolation although the
density-density correlations are short range. Therefore, it would be
very interesting to consider an analytical calculation for this model.

\begin{acknowledgments}
  The authors thanks Malte Schröder and Johannes Zierenberg
  for uselful discussion.
  The simulations were performed at the
  the HPC cluster ROSA, located at the University of Oldenburg
  (Germany) and
    funded by the DFG through its Major Research Instrumentation Program
    (INST 184/225-1 FUGG) and the Ministry of
    Science and Culture (MWK) of the
    Lower Saxony State.
\end{acknowledgments}

\bibliography{lit}

\end{document}